\documentclass{article}

\usepackage{arxiv}

\usepackage[utf8]{inputenc}
\usepackage[T1]{fontenc}
\usepackage[colorlinks,bookmarksopen,bookmarksnumbered,citecolor=blue,urlcolor=blue]{hyperref}
\usepackage{url}            
\usepackage{booktabs}       
\usepackage{amsfonts}       
\usepackage{nicefrac}       
\usepackage{microtype}      
\usepackage[square,numbers]{natbib}
\usepackage{doi}
\usepackage{amsmath}
\usepackage{amssymb}
\usepackage{cleveref}
\usepackage{bm}
\usepackage{physics}
\usepackage{graphicx}
\usepackage{tikz}

\newcommand{\ipc}[2]{\langle #1,\, #2\rangle}
\newcommand{\ipcf}[2]{\ipc{#1}{#2}_F}
\renewcommand{\norm}[1]{\|#1\|}
\newcommand{\normf}[1]{\norm{#1}_F}
\renewcommand{\ker}[2]{\kappa\left( #1,\, #2\right)}
\newcommand{\kerp}[3]{\kappa_{#1}\left( #2,\, #3\right)}
\newcommand{\bve}{\bm{\varepsilon}}

\title{Predicting Open-Hole Laminates Failure Using Support Vector Machines With Classical and Quantum Kernels}

\date{}

\newif\ifuniqueAffiliation
\uniqueAffiliationtrue

\ifuniqueAffiliation
\author{Giorgio Tosti Balducci\\
	Aerospace Structure and Materials\\
	Delft University of Technology\\
	Kluyverweg 1, 2629HS,\\
        Delft, The Netherlands\\
	\And
        Boyang Chen\thanks{Corresponding author. Email: \texttt{b.chen-2@tudelft.nl}}\\
	Aerospace Structure and Materials\\
	Delft University of Technology\\
	Kluyverweg 1, 2629HS,\\
        Delft, The Netherlands\\
        \And
        Matthias M\"{o}ller \\
        Applied Mathematics \\
	Delft University of Technology\\
        Mekelweg 4, 2628CD, \\
        Delft, The Netherlands\\
	\And
        Marc Gerritsma\\
	Flow Physics and Technology\\
	Delft University of Technology\\
	Kluyverweg 1, 2629HS,\\
        Delft, The Netherlands\\
	\And
        Roeland De Breuker\\
	Aerospace Structure and Materials\\
	Delft University of Technology\\
	Kluyverweg 1, 2629HS,\\
        Delft, The Netherlands\\
}
\else
\fi

\renewcommand{\shorttitle}{Open-Hole Composite Failure with Classical and Quantum SVM}

\hypersetup{
pdftitle={Predicting Open-Hole Laminates Failure Using Support Vector Machines With Classical and Quantum Kernels},
pdfsubject={cs.CE, cs.AI, quant-ph},
pdfauthor={Giorgio Tosti Balducci, Boyang Chen, Matthias Möller, Marc Gerritsma, Roeland de Breuker},
pdfkeywords={Composites, Support Vector Machines, Quantum Machine Learning},
}

\begin{document}
    \maketitle
    \begin{abstract}
        Modeling open hole failure of composites is a complex task, consisting in a highly nonlinear response with interacting failure modes. Numerical modeling of this pheonomenon has traditionally been based on the finite element method, but requires to tradeoff between high fidelity and computational cost. To mitigate this shortcoming, recent work has leveraged machine learning to predict the strength of open hole composite specimens. Here, we also propose using data-based models but to tackle open hole composite failure from a classification point of view. More specifically, we show how to train surrogate models to learn the ultimate failure envelope of an open hole composite plate under in-plane loading. To achieve this, we solve the classification problem via support vector machine (SVM) and test different classifiers by changing the SVM kernel function. The flexibility of kernel-based SVM also allows us to integrate the recently developed quantum kernels in our algorithm and compare them with the standard radial basis function (RBF) kernel. Finally, thanks to kernel-target alignment optimization, we tune the free parameters of all kernels to best separate safe and failure-inducing loading states. The results show classification accuracies higher than 90\% for RBF, especially after alignment, followed closely by the quantum kernel classifiers.

    \end{abstract}

    \keywords{Composites \and Support Vector Machines \and Quantum Machine Learning}

    \section{Introduction}
    \label{sec:intro}
        Modern aviation industry makes wide use of composite materials, thanks to their lightweight and favorable mechanical properties. Frequently, aeronautical structural elements are often not textbook flat composite panels, but tailored components with complex mechanical responses. For instance, composite panels often show cutouts in order to allow fastening or lightening the structure or even for allowing the passage of wiring or cables. However, the presence of holes in a composite plate induces stress concentrations that can initiate damage which can propagate into intricate failure mechanisms involving different modes.

        Models for open hole composite failure have developed in different directions.
        On the one hand, semi-empirical models were proposed to predict the allowables of these structures, such as ultimate stength, and their statistical distribution with respect to hole geometry, loading conditions, stacking sequence, ply thickness, etc. Early attempts required experimental properties from testing both the unnotched and notched laminate \cite{Whitney1974}, while later models removed the need of directly testing the open hole laminate \cite{Camanho2012, Catalanotti2021} or just required the ply properties \cite{Furtado2017}. Despite being fast to evaluate and suitable for preliminary design, semi-empirical models can make large errors when extensive delaminations propagate from the notch, as it happens with ply-scaled laminates.

        Finite Element (FE) simulations allow for improved modeling of open hole laminates failure. Open hole tension (OHT) has been extensively studied numerically both for capturing the in-plane \cite{Camanho2007} and thickness size effects \cite{Hallett2009,vanderMeer2011,Chen2013} on the ultimate strength and for reproducing the different failure modes and their interactions \cite{vanderMeer2010,Chen2016} with increasing detail. Furthermore, FE simulations managed to quite accurately predict open-hole compression (OHC), even though still struggling to predict the precise kink band formation\cite{Soutis1991,su2015progressive,HIGUCHI2021106300}. However, the accuracy offered by FE models generally comes at the price of high computational costs, possibly making them unfeasible when many design iterations are required.

        Therefore, there is a practical need for computationally efficient yet accurate models that can simulate open hole composite laminates. A possibility is offerred by machine learning surrogates, which have been employed in composite design and optimization \cite{Bisagni2002,Bessa2018,Zhang2022Apr}, constitutive law modeling and multiscale analyses (see \cite{Liu2021} for a comprehensive review) and damage characterisation \cite{Zobeiry2020,Reiner2021Oct}. Concerning open-hole composite failure, Furtado \emph{et al.} proposed a methodology to define allowables using four different machine learning models \cite{Furtado2021}. Their methodology was applied to open-hole tensile strength prediction for different dimensions, layups and material properties. While their methods are demonstrated on data generated analytically \cite{Furtado2017}, the authors suggest using high fidelity finite element analyses for training, potentially providing accurate data-based models. 

        Similarly, in this work we propose a machine learning surrogate for open hole composites, which is accurate and efficient in inference. Differently from \cite{Furtado2021} however, the approach we suggest is not to have a fast allowables generator, but a classifier for ultimate failure of open hole composite laminates. More precisely, our trained model takes a loading state as input, such as the far field homogenized plane strain components and returns a binary valuable ($\pm1$) as output, depending on whether the load applied is lower or higher than the notched laminate strength. In this sense, the surrogate acts as a data-based generalized failure criterion which predicts at the structural component level, rather than at the material level.

        This paper also aims at comparing classical and quantum computation for a classification problem in composite mechanics. To do this, we train the machine learning surrogate using \emph{kernel}-based support vector machines (SVMs) \cite{Cortes1995Sep}, where the kernel function can be computed both in classical and quantum logic. As it will be clear in the next sections, quantum computation offers a way to encode information into exponentially large Hilbert spaces and to define an inner product in this spaces, effectively generating a kernel. This allows to explore the generalization potential of quantum machine learning, while leaving the SVM optimization to well-established classical quadratic optimization algorithms.

        The rest of the paper is structured as follows. \Cref{sec:ml_problem} describes the machine learning problem, by defining the input, the data sampling strategy and the labeling criterion. \Cref{sec:methodology} briefly introduces the SVM dual problem, the Radial Basis Functions (RBF) kernels and the quantum kernels. More details about these methods are available in the appendices following the main body of the manuscript. Finally, \Cref{sec:results} presents the classification results  for all kernels and \Cref{sec:conclusion} outlines conclusions and future work.
        
        All data and code used in this work are made publicly available (see \cite{boyang_chen_2022_7409612}, \cite{TostiBalducci2023Code} respectively).

    \section{Machine learning problem}
    \label{sec:ml_problem}

        \begin{figure}[tbp]
            \centering
            \includegraphics[width=0.5\textwidth]{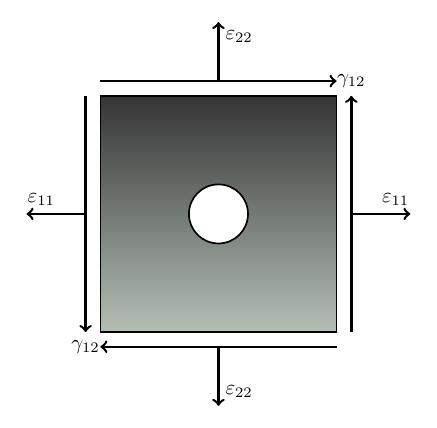}
            \caption{Geometry of the open hole composite plate and homogenous strain loads components.}
            \label{fig:geometry}
        \end{figure}

        Our method was applied to predict failure of an open hole composite specimen similar in geometry and material properties to the one experimentally tested in \cite{Green2007}. The specimen was modeled and meshed with the Abaqus finite element code \cite{abaqus2022} and it was loaded with different combinations of axial and shear strains and constrained with periodic boundary conditions. All the details of the specimen properties and of the finite element analyses are left to \Cref{sec:appa}.

        The input of our surrogate models are homogenized far field strains $\bm{\varepsilon}=\left[ \varepsilon_{11},\, \varepsilon_{22},\, \gamma_{12} \right]^\top$, which derive from enforcing periodic boundary conditions on opposite faces of the plate. The displacements of the left/right and top/bottom faces respectively can be linked through some reference degrees of freedom
        \begin{equation}
            \begin{aligned}
                U_1 &= u_1^R - u_1^L\\
                U_2 &= u_2^T - u_2^B\\
                U_3 &= u_2^R - u_2^L\\
                U_4 &= u_1^T - u_1^B,
            \end{aligned}
            \label{eq:rp_displacements}
        \end{equation}
        where directions $1$ and $2$ are the horizontal and vertical directions in \Cref{fig:geometry}. The homogenized strains are then obtained as
        \begin{equation}
            \begin{aligned}
                \varepsilon_{11} &= \frac{U_1}{D_1}\\
                \varepsilon_{22} &= \frac{U_2}{D_2}\\
                \gamma_{12} &= \frac{U_3}{D_1} + \frac{U_4}{D_2},
            \end{aligned}
            \label{eq:homogenized_strains}
        \end{equation}
        where $D_1$ and $D_2$ are the planar dimensions of the plate.
        
        \begin{figure}[tbp]
            \centering
            \includegraphics[width=\textwidth]{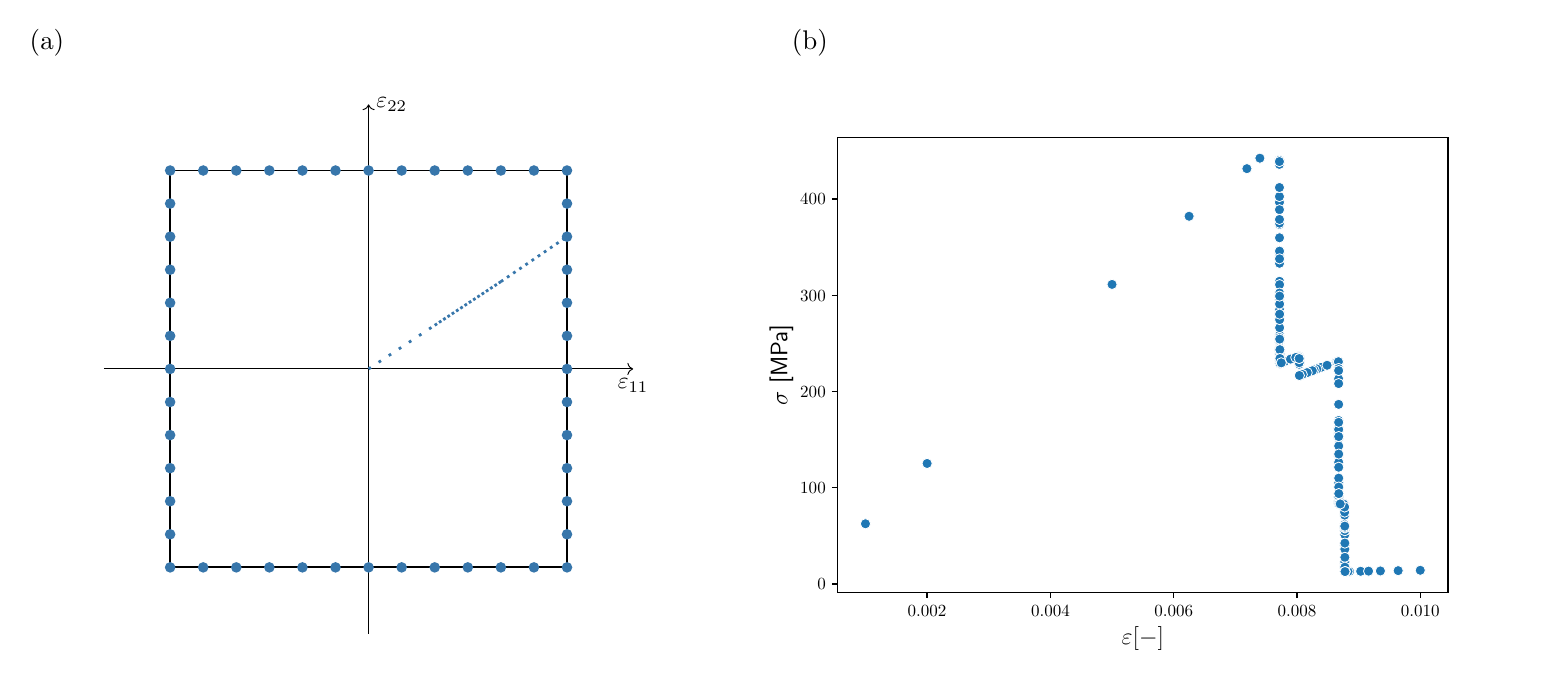}
            \caption{(a) Radial strategy for sampling the far-field strain space. The bigger dots on the edge of the $\left( \varepsilon_{11}, \varepsilon_{22} \right)$ volume represent the final applied load in different nonlinear FE incremental-iterative analyses. The arrow represents a loading path with the load increments unevenly distributed. (b) Load-displacement curve corresponding to the loading path in (a).}
            \label{fig:sampling}
        \end{figure}

        As mentioned, the input space was sampled through nonlinear incremental-iterative finite element analyses. \Cref{fig:sampling} illustrates the sampling strategy used in this work in the simplified case of two-dimensional input. We refer to this technique as \emph{radial sampling}, due to the fact that the design of experiments (DoE) does not directly affect all the points in this input space, but only the ones on the boundary. On the other hand, all the intermediate points are generated internally by the FE solver and they correspond to the homogenized strain values at every time increment. The user maintains control of the inner samples values, by the choice of initial, minimum and maximum time steps. For this work, we chose the sampling space to be the hypercube $\left[ -10^{-2},\, 10^{-2} \right]^{\otimes 3}$ in $\mathbb{R}^3$, meaning that all three components of the applied strains vector have the same bounds.

        \subsection{Labeling criterion}
        \label{sec:pp}
        Each strain sample was assigned a label based on an ultimate failure criterion. In particular, we defined failure by the loss of stiffness of the laminate for given a user-defined threshold.

        From the results of the FE analyses with periodic boundary conditions, one obtains the reaction forces $F_1$, $F_2$, $F_3$ and $F_4$ conjugate to the degrees of freedom in \Cref{eq:rp_displacements}. These provide the homogenized stresses, which can then be derived via the Hill-Mandel principle of energy balance as
        \begin{equation}
            \begin{aligned}
                \sigma_{11} &= \frac{F_1}{tD_2}\\
                \sigma_{22} &= \frac{F_2}{tD_1}\\
                \sigma_{12} &= \frac{F_3U_3+F_4U_4}{\gamma_{12}tD_1D_2},
            \end{aligned}
            \label{eq:homogenized_stresses}
        \end{equation}
        where $t$ is the thickness of the plate.

        The laminate stiffness in the two axial directions and in shear can thus be defined at every timestep $t$ as
        \begin{equation}
            \begin{aligned}
                E_1^{(t)} &= \frac{\sigma_{11}^{(t)}}{\varepsilon_{11}^{(t)}}\\
                E_2^{(t)} &= \frac{\sigma_{22}^{(t)}}{\varepsilon_{22}^{(t)}}\\
                G_{12}^{(t)} &= \frac{\sigma_{12}^{(t)}}{\varepsilon_{12}^{(t)}}.
            \end{aligned}
            \label{eq:laminate_stiffness}
        \end{equation}

        The stiffness degradation $d_S$ is defined as the minimum ratio between the instantaneous stiffness and the corresponding stiffness measure in the linear elastic region,
        \begin{equation}
            d_S^{(t)}=\min\left\{ \frac{E_1^{(t)}}{E_1^{(0)}},\, \frac{E_2^{(t)}}{E_2^{(0)}},\,\frac{G_{12}^{(t)}}{G_{12}^{(0)}} \right\}
            \label{eq:stiffness_degradation}
        \end{equation}

        Therefore, given $M$ the total number of samples, every sample $\bve^{(m)}$ ($m=1,\dots, M$) is assinged a label $y^{(m)}=-1$ if $d^{(m)}<\bar{d}_S$ and $y^{(m)}=+1$ otherwise. 

    \section{Methodology}
    \label{sec:methodology}
        \begin{figure}[tbp]
            \centering
            \includegraphics[width=\textwidth]{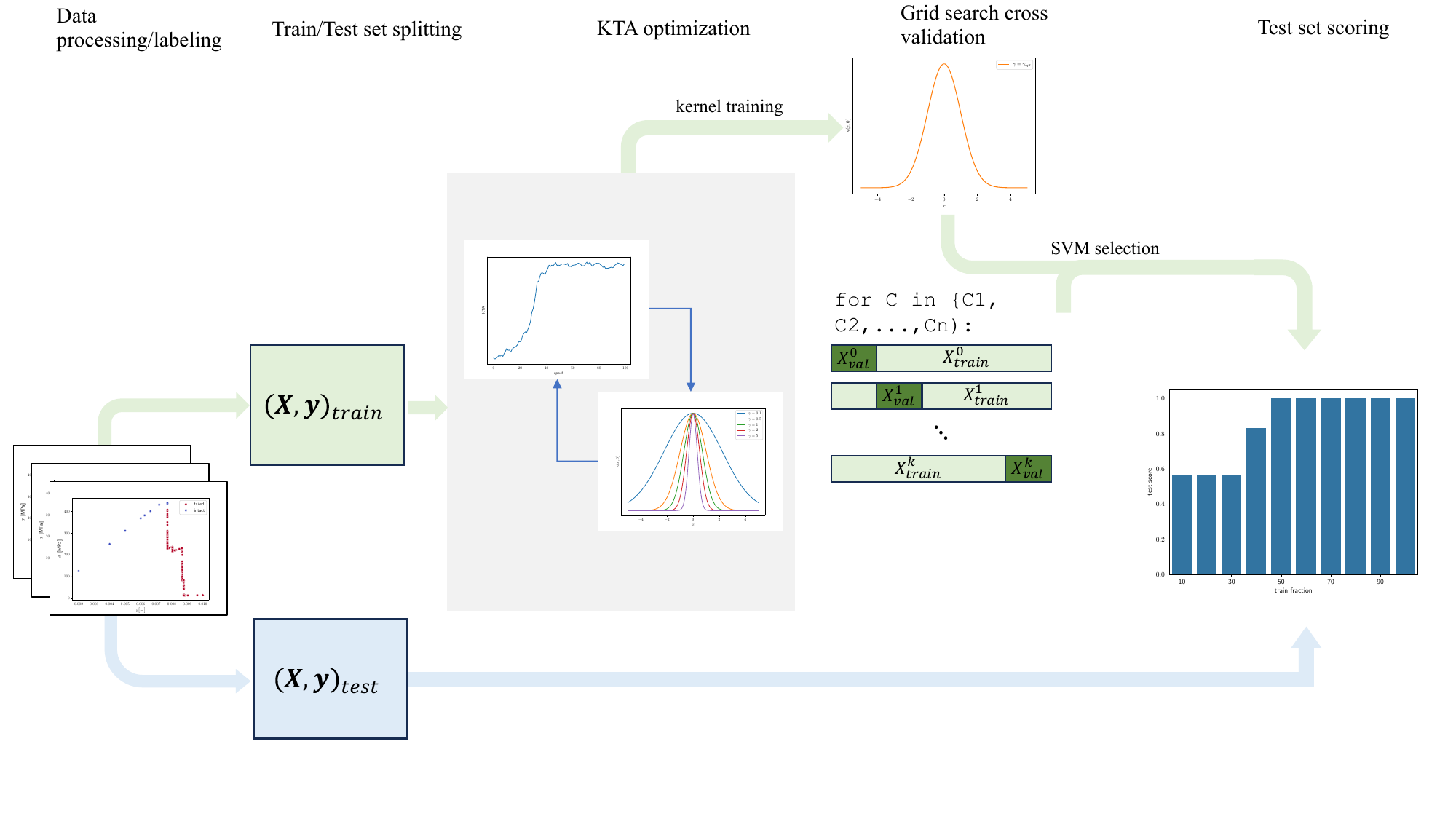}
            \caption{The methodology used in this work. The dataset is generated by nonlinear finite element analyses, then labeled and split into training and testing sets. The training set is used first to train the kernel, by optimizing the kernel-target alignment (KTA), then in a grid search cross-validation to find the best slack penalty $C$ of the SVM. With all the hyperparameters fixed, the SVM is trained for increasing dataset sizes and the classification accuracy is evaluated on the testing set.}
            \label{fig:methodology}
        \end{figure}
        As already mentioned, we solve the ultimate failure binary classification problem using the SVM algorithm \cite{Cortes1995Sep}. This consists in the following quadratic optimization problem in dual form 
        \begin{equation}
            \begin{aligned}
                \max_{\bm{\alpha}}&\quad \sum_{m=1}^{M}\alpha^{(m)}-\frac{1}{2}\sum_{m,m^\prime=1}^{M}y^{(m)}y^{(m^\prime)}\alpha^{(m)}\alpha^{(m^\prime)}\ker{\bm{x}^{(m)}}{\bm{x}^{(m^\prime)}}\\
                \mathrm{s.t.}&\quad 0\leq\alpha^{(m)}\leq C,\quad m=1,\,\dots,\,M\\
                &\quad\sum_{m=1}^{M}\alpha^{(m)}y^{(m)}=0,
            \label{eq:met_svm_dual}
            \end{aligned}
        \end{equation}
        where $\bm{x}^{(m)} = \bm{\varepsilon}^{(m)}$, $y^{(m)}=\pm 1$ are the labels, respectively non-failed and failed, $\alpha^{(m)}$ are the Lagrange multipliers and $C$ is the slack penalty. The kernel function $\ker{\cdot}{\cdot}$ is a similarity metric between two samples in a higher-dimensional feature space. More details on the SVM algorithm are left to \Cref{sec:appb}.
        
        The performance of the dual SVM depends on the choice of its hyperparameters, namely the kernel function $\kappa$ and slack penalty $C$. To restrict the search space, the kernel function is generally parametrized via one or more parameters $\bm{ \theta }$ and the standard practice is to perform a grid-search cross-validation procedure in the $\left(\bm { \theta }, C \right)$ space. In this work, we use instead a mixed procedure, where the kernel function is determined by optimizing the kernel-target alignment (KTA) \cite{wang2015KtaReview} and the slack penalty is found by grid search cross-validation. The overall methodology is illustrated in \Cref{fig:methodology}, where we refer to the two steps as \emph{kernel training} and \emph{SVM selection}. Once the SVM has been fully determined, it can be trained by solving \Cref{eq:met_svm_dual} and its learning ability can be measured as the accuracy on unseen test data, for different training dataset sizes.

        We compare one classical and two quantum kernels. The classical kernel is the radial basis function (RBF) kernel, defined as
        \begin{equation}
            \kerp{\text{RBF}}{\bm{x}^{(m)}}{\bm{x}^{(m^\prime)}} = \exp\left( -\gamma\norm{\bm{x}^{(m)}-\bm{x}^{(m^\prime)}}^2\right).
            \label{eq:met_rbf_kernel}
        \end{equation}
        RBF is a powerful kernel which corresponds to a feature map in an infinite-dimensional feature space \cite{Schoelkopf2019}. It induces a Gaussian similarity function, whose width is controlled by the hyperparameter $\gamma$. 
        
        On the other hand, the quantum kernel is defined via a \emph{quantum embedding}, which is constructed via data-depending unitary transformations $U\left( \bm{x} \right)$ that prepare the quantum state
        \begin{equation}
            \ket{\psi\left( \bm{x} \right)} = U\left( \bm{x} \right)\ket{0}.
            \label{eq:met_quantum_embedded_state}
        \end{equation}
        Given two samples $\bm{x}^{(m)}$ and $\bm{x}^{(m^\prime)}$, the quantum kernel is simply the inner product
        \begin{equation}
            \kerp{\text{Q}}{\bm{x}^{(m)}}{\bm{x}^{(m^\prime)}} = \abs{\ipc{\psi\left( \bm{x}^{(m)} \right)}{\psi\left( \bm{x}^{(m^\prime)} \right)}}^2.
            \label{eq:met_quantum_kernel}
        \end{equation}

\begin{figure}[tbp]
            \centering
            \includegraphics[width=\textwidth]{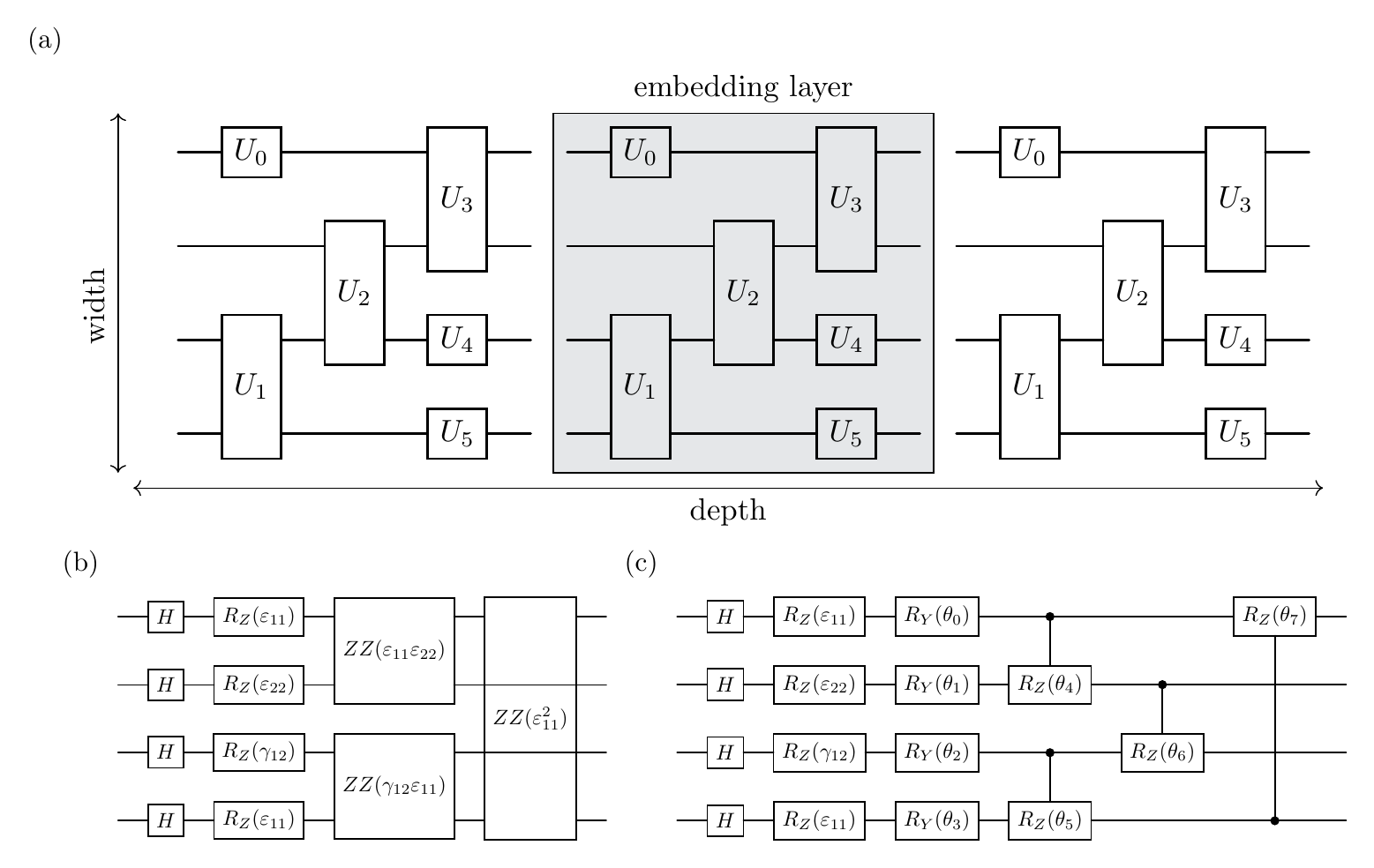}
            \caption{Quantum embeddings. (a) Generic quantum embedding made of one- and two-qubit gates. The \emph{width} of the embedding is the number of qubits, while the \emph{depth} is the number of layers, which is a minimal block of gates. (b) IQP quantum embedding layer, which is parameter free, but encodes products of features as $ZZ$ interactions. (c) HE2 quantum embedding layer, parametrized by $\left[ \theta_0, \theta_1, \dots, \theta_{P-1} \right]$. }
            \label{fig:met_qc_embeddings}
        \end{figure}

        \Cref{fig:met_qc_embeddings} shows the generic quantum embedding and the two specific ones used in this work, which are the hardware efficient embedding (HE2) \cite{Hubregtsen2021} and the instantaneous quantum polynomial (IQP) \cite{Kyriienko2022} one. To have more expressive feature mapping, either the \emph{width} or the \emph{depth} of the quantum embedding can be increased. The first one is the number of qubits, which can be even higher than the number of features in the dataset, by cyclically re-encoding the features to generate a highly nonlinear and potentially better separable feature space. Meanwhile, the embedding's depth can be increased by repeating a base data-encoding block, such as IQP and HE2. Even in this case, re-encoding of the features may lead to a higher expressivity of the overall feature map \cite{Schuld2020}. For a short summary of relevant quantum computing concepts, we refer the reader to \Cref{sec:appc}.

        \section{Results}
        \label{sec:results}
        We tested our machine learning models on a dataset of 1960 labelled strain vectors $\bm{\varepsilon}^{(m)}$, which we obtained by uniformly sampling the homogeneous strain/stress pairs from the FE simulations of the open-hole composite specimen. The input homogeneous strains in both normal and shear directions were varied between $10^{-4}$ and $10^4$ microstrains and a stiffness degradation threshold of 0.9 was used to discriminate non-failed and failed loading states.

        Both classical- and quantum-kernel SVMs were implemented using different Python libraries. We used PyTorch for training the RBF kernel and PennyLane for the quantum kernels. These libraries implement automatic differentiation (AD), which allows to optimize the KTA with gradient-based methods. We also used JAX together with PennyLane to just-in-time compile the quantum kernel functions. Concerning the classification problem, we employed the SVM and grid-search cross validation routines available from the Scikit-Learn Python package.

        \begin{figure}[tbp]
            \centering
            \includegraphics[width=0.6\textwidth]{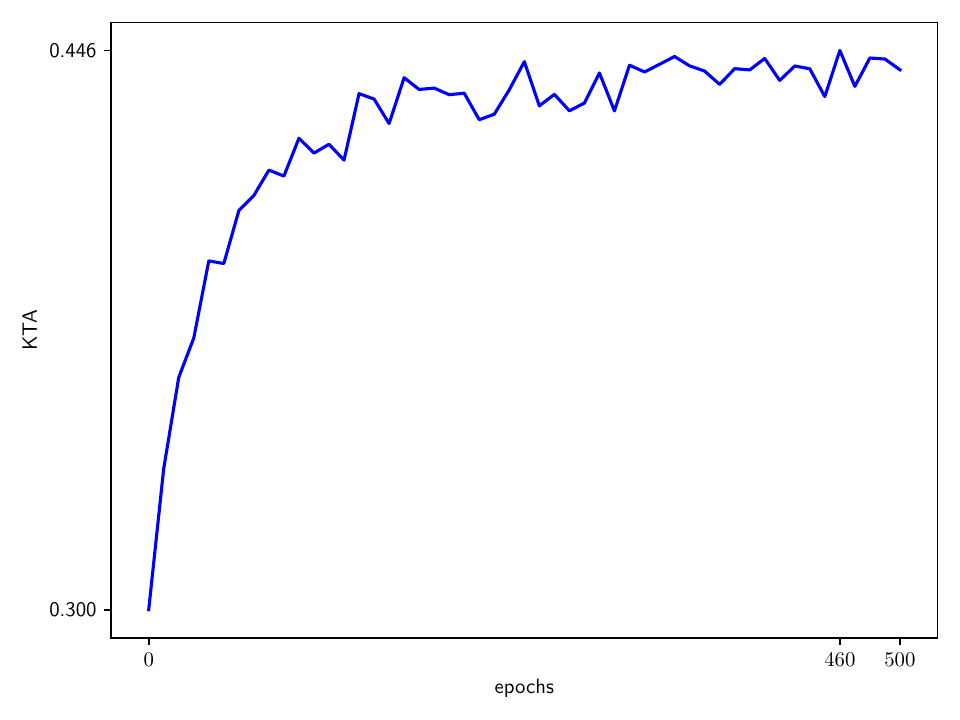}
            \caption{Training history of the RBF kernel's KTA.}
            \label{fig:res_rbf_kta_opt}
        \end{figure}

        \begin{figure}[tbp]
            \centering
            \includegraphics[width=0.8\textwidth]{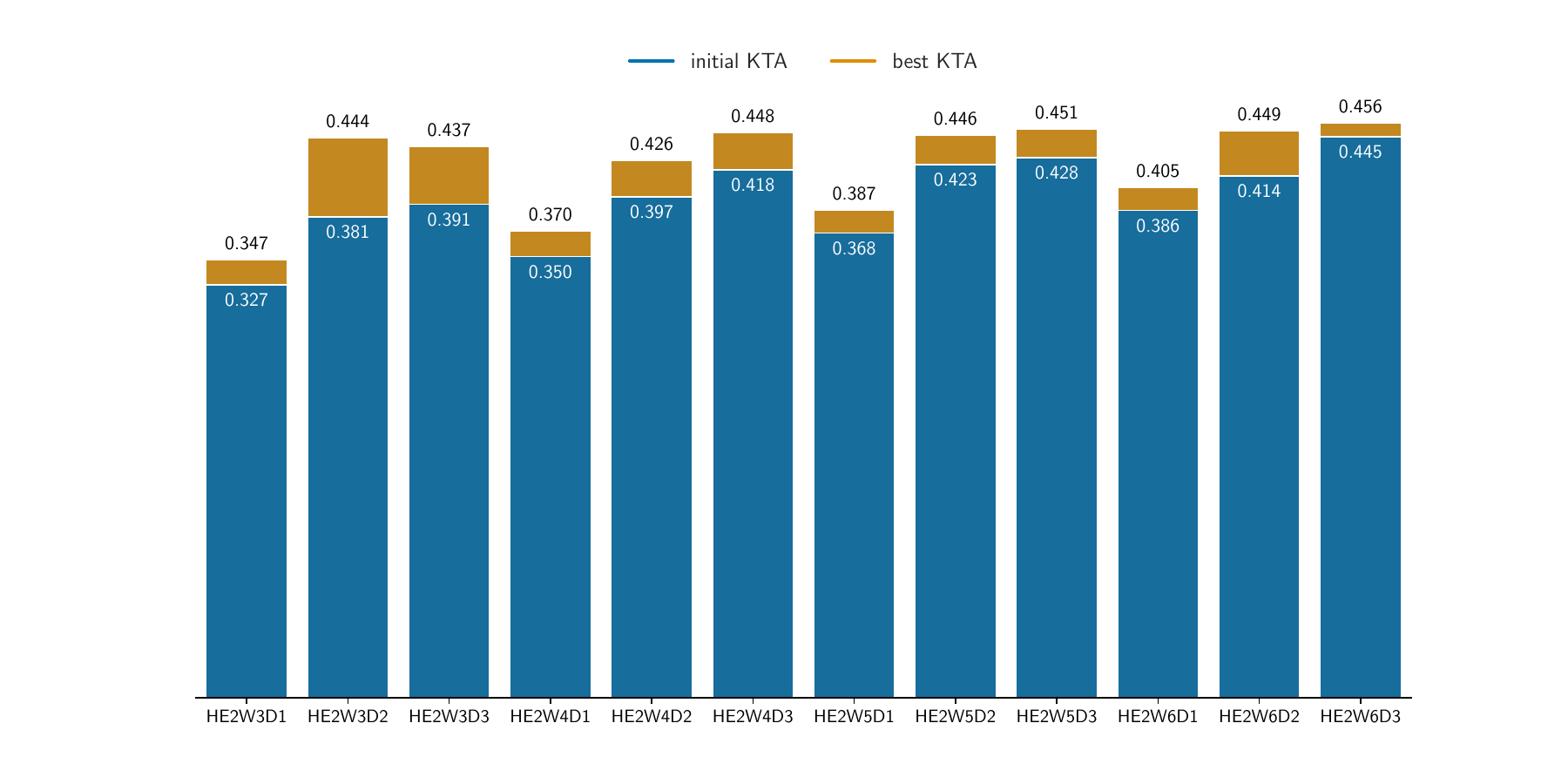}
            \caption{KTA of quantum kernels before and after training for 9 different quantum embeddings. The basic layer for all embeddings is the HE2, which has trainable parameters. The tag WXDY indicates width and depth of the embedding.}
            \label{fig:res_quantum_kta_opt}
        \end{figure}

        The KTA of both RBF and quantum kernels was maximized using stochastic gradient descent and Adam parameters update \cite{Kingma2014}. \Cref{fig:res_rbf_kta_opt} shows the kernel alignment training of the RBF kernel. \Cref{fig:res_quantum_kta_opt} presents instead the KTAs before and after training for nine different quantum kernels with HE2 embedding. It can be seen that increasing width and depth of these kernels generally improves their KTA. A higher number of qubits means that the strain features are mapped in a higher dimensional space, which can favor separability of the classes. On the other hand, increasing the depth benefits the kernel alignment, since it results in more expressive feature maps. Also, every additional layer of the HE2 embedding doubles the number of free parameters, explaining why optimization of deeper kernels mostly leads to higher gains in KTA. However, the advantage of increasing these quantum encoding resources does not scale uniformly. Already with 6 qubits and 3 HE2 layers, the optimization only modestly improves the KTA, likely due to the vanishing KTA gradients \cite{McClean2018}.
        
        \begin{figure}[htbp]
            \centering
            \includegraphics[width=0.7\textwidth]{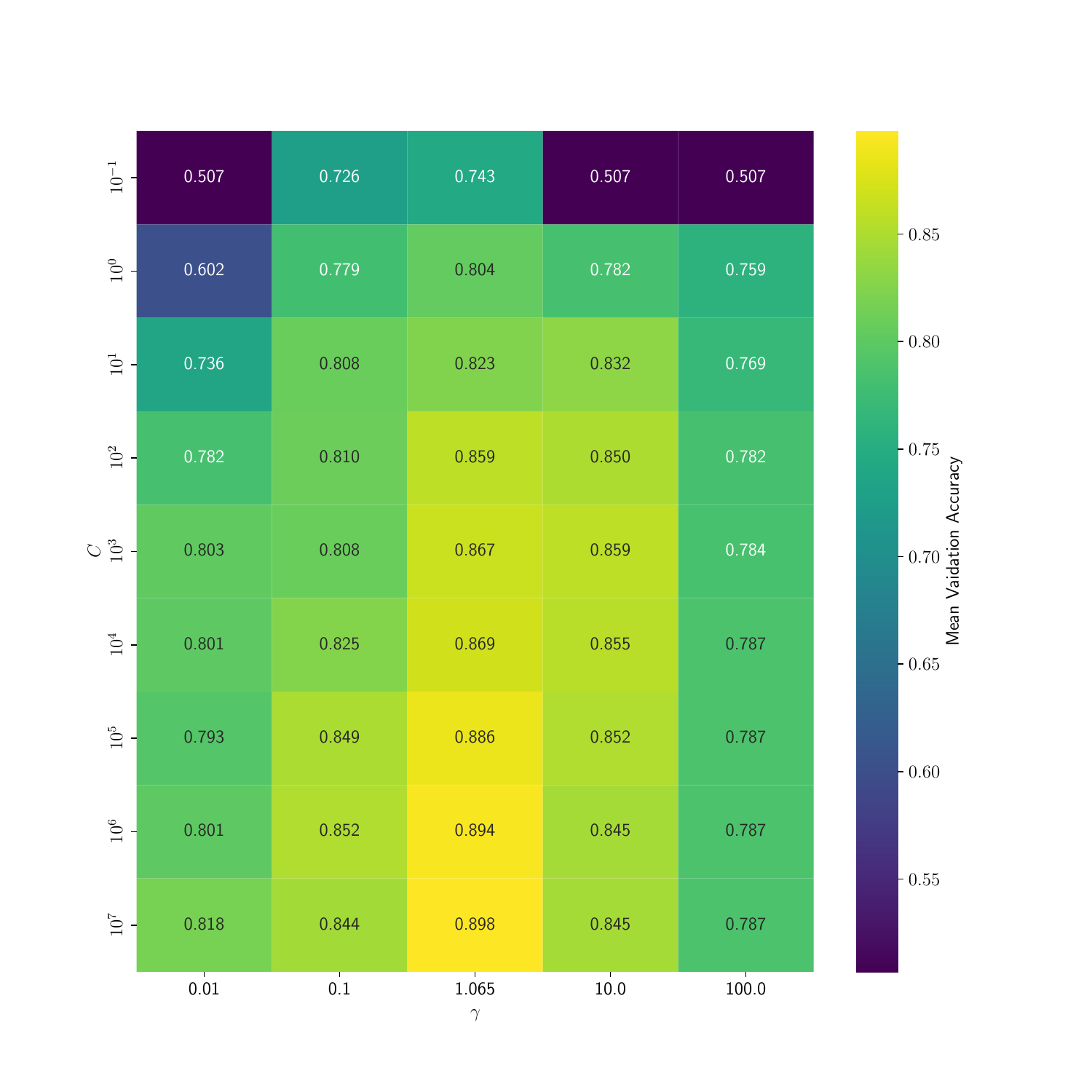}
            \caption{Grid-search cross validation for the RBF kernel. $\gamma=1.065$ corresponds to the kernel with highest kernel-target alignment.}
            \label{fig:res_gscv_rbf}
        \end{figure}

        To find the hyperparameter $C$ that guarantees the highest off-training accuracy of the SVM algorithm, we used grid-search cross validation for the kernels considered. The validation accuracy values are reported in \Cref{fig:res_gscv_rbf} for multiple values of $C$ and $\gamma$. We observe that kernels with $\gamma\leq 10$ achieve the higest scores, with the highest-KTA $\gamma$ scoring first for the whole range of $C$ values. Furthermore, the accuracy of the maximally-aligned RBF kernel increases monotonically with $C$, which suggests the usefulness of maximizing the KTA, but also that the class boundary in this feature space is densely populated and still requires a tight margin. 

        \begin{figure}[htbp]
            \centering
            \includegraphics[width=\textwidth]{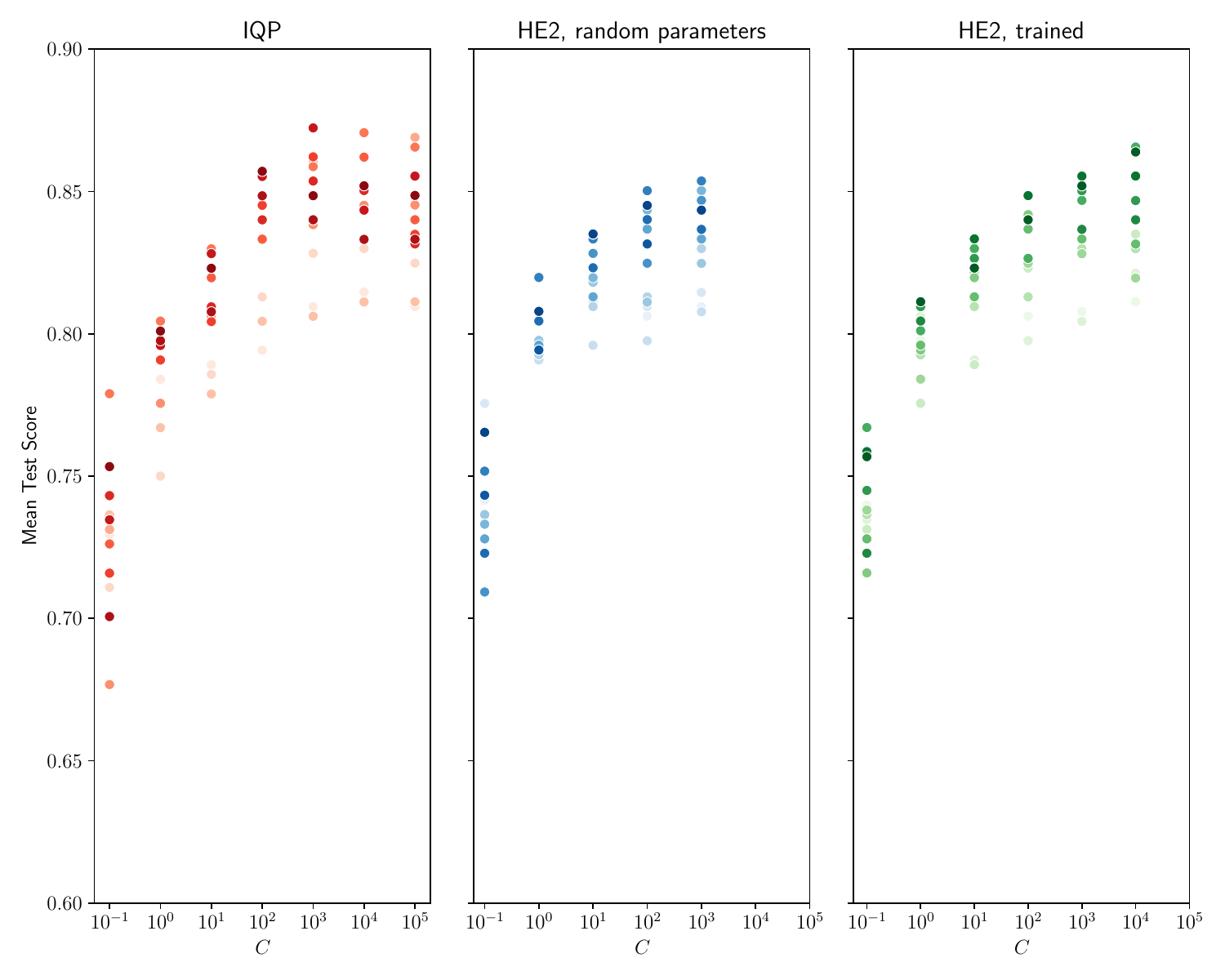}
            \caption{Summary of the grid-search cross validation results for different quantum kernels. The three figures correspond respectively to the IQP embeddding, the HE2 kernel with untrained parameter and HE2 with trained parameters. Different shades of the same color correspond to different depths and widths. From lighter to darker, the points correspond to embeddings of increasing widths and of increasing depth per fixed width.}
            \label{fig:res_gscv_quantum}
        \end{figure}

        The same analysis was performed for all the quantum kernels considered, where we wanted to take into account the effect on accuracy of different embeddings and of maximizing the kernel-target alignment. The results are reported in \Cref{fig:res_gscv_quantum}, which shows accuracies roughly between 67\% and 87\% for all embeddings with different values of $C$. Except for IQP case, increasing $C$ leads to higher accuracies, hinting to the the need of a tight bound when mapping with these embeddings, similar to the RBF kernel. Unfortunately, for high values of $C$, the optimization of the dual SVM failed to converge for some of the quantum kernels, likely due to numerical ill-conditioning. This presumably prevented the quantum kernel classifiers from even better separating failed instances, as suggested by the monotonic increasing test accuracies with $C$, at least for the HE2 kernels. Furthermore, \Cref{fig:res_gscv_quantum} shows that the scores improve when more embedding resources (number of qubits and layers) are added, especially in the case of KTA-optimized HE2 kernel.

        \begin{figure}[tbp]
            \centering
            \includegraphics[width=\textwidth]{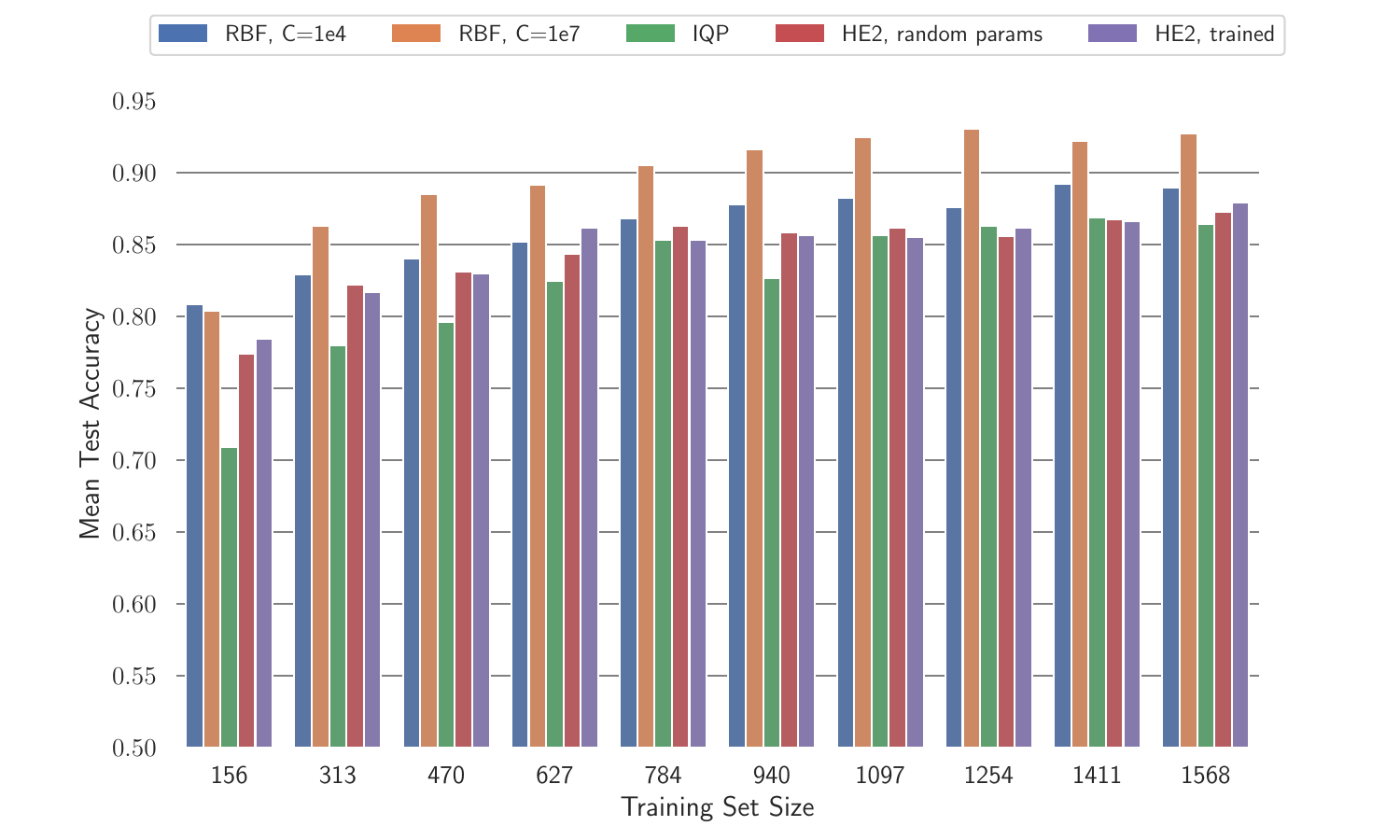}
            \caption{Test data classification accuracy of classical and quantum kernels for increasing training set size. The RBF kernel-SVMs were trained with  both $C=10^7$ and $C=10^4$, as the latter is the highest value of $C$ for which the quantum-kernel SVMs could still be fitted. Quantum-kernel classifiers were trained instead with the embedding arhitecture and $C$ value ensuring highest accuracy during grid-search cross validation.}
            \label{fig:res_test_acc}
        \end{figure}

        Classical and quantum kernels are finally compared in \Cref{fig:res_test_acc}, which shows how 5 different models classify a test set of strain loading data when fitted on progressively larger training sets. A similar comparison on additional classification metrics can be found in \Cref{sec:appd}. The RBF kernel achieves 80\% accuracy with just 10\% the total training set size, and with $C=10^7$ it reaches over 90\% with just half the training points. In comparison, all quantum kernel classifiers are at least 5\% less accurate than the best RBF-kernel SVM. However, especially for HE2 embeddings, the scores are similar to the $C=10^4$ RBF case, suggesting that RBF and HE2 kernels separate the non-failed and failed classes to a similar extent. Changing the embedding from HE2 to IQP, there is  a drop in accuracy for small training set sizes, while the performance is similar when more than half the training set is used. On the other hand, the effect of training the kernel is less visible at this stage, reflecting the fact that the accuracies obtained during grid-search cross validation are alike for untrained and trained HE2.

        \section{Conclusion}
        \label{sec:conclusion}
        In this paper, we proposed a methodology to build a binary classifier from finite element analyses data for the particular case of an open hole composite specimen. We studied the case of in-plane strain loading of the specimen where the objective is to correctly label strain combinations that lead to ultimate failure.

        From a design of experiment point of view, we demonstrated a radial sampling strategy technique, where the choice of which simulations to make to cover the input space takes into account the incremental-iterative nature of the nonlinear FE method. We then proposed a labelling criterion of homoegenized strain-stress pairs based on residual in-plane stiffness.

        For classification of the labelled data, we used the kernel-based SVMs, which also allowed us to compare the performance of the recently proposed quantum kernels against the more traditional RBF. Furthermore, we employed kernel-target alignment to improve class separbility of both RBF and the HE2 embedding kernel.

        For all the kernel examined, the corresponding SVMs separate non-failed and failed loading states with good accuracy. The RBF-based model classify more accurately than its quantum counterparts, although this likely happens due to numerical ill-conditioning in the current quantum SVM implementation. These numerical issues can likely be fixed by studying the dual SVM problem for the problematic instances, which will be the subject of future work.

        Regarding kernel alignment, optimizing the KTA is shown to be powerful for RBF, since the SVM for the trained kernel outperforms the other RBF-based models in terms of accuracy. Aligning quantum kernels for this dataset also helps them to better separate the two classes, but for simple architectures the improvement is moderate, while more complex embeddings only reach the scores of the more simple ones after they have been aligned. Furthermore, one should remember that optimizing quantum kernels is almost always more computationally involved than for RBF, as the formers can have highly parametrized embeddings, while RBF is completely defined by the single parameter $\gamma$.

        Extensions of this work can go in many directions. From the point of view of the problem, it would be interesting to increase the number of degrees of freedom, by allowing the notch radius or the lamination sequence to also change. The latter could be written in terms of lamination parameters \cite{Setoodeh2006} to have a continuos representation.

        In terms of algorithms, both classical and quantum kernels can be explored further. RBF is the most popular choice for classical kernels, but certainly not the only one. Due to Mercer's condition, any function which defines a positive semi-definite kernel matrix is a valid kernel function \cite{Schoelkopf2019}. Obviously, the design space is vast, but automated procedures help reduce the search for instance by exploring combinations of only a fixed set of standard kernel functions.

        On the other hand, the freedom of designing and parametrizing quantum embedding circuit also makes the choice of a quantum kernel nontrivial. Within the limits of classical simulation of quantum circuits, one could experiment with increasing number of qubits or different layering strategies, for instance the one proposed in \cite{Miroszewski2023} for the task of satellite image classification. From an optimization point of view, a recent technique has been proposed to maximize the quantum kernel alignment KTA and solving the SVM in a single optmization loop \cite{Gentinetta2023}, which would of course greatly reduce the computational cost. Nevertheless, to truly understand a potential competitiveness of quantum kernels, it is probably most important to remove layers of simulation and study the effects of statistical and hardware noise on SVM convergence and accuracy.

    \appendix

    \section{Open hole specimen features and finite element model details}
    \label{sec:appa}

    \subsection{Geometry and material properties}
        \begin{figure}[tbp]
            \centering
            \includegraphics[width=0.7\textwidth]{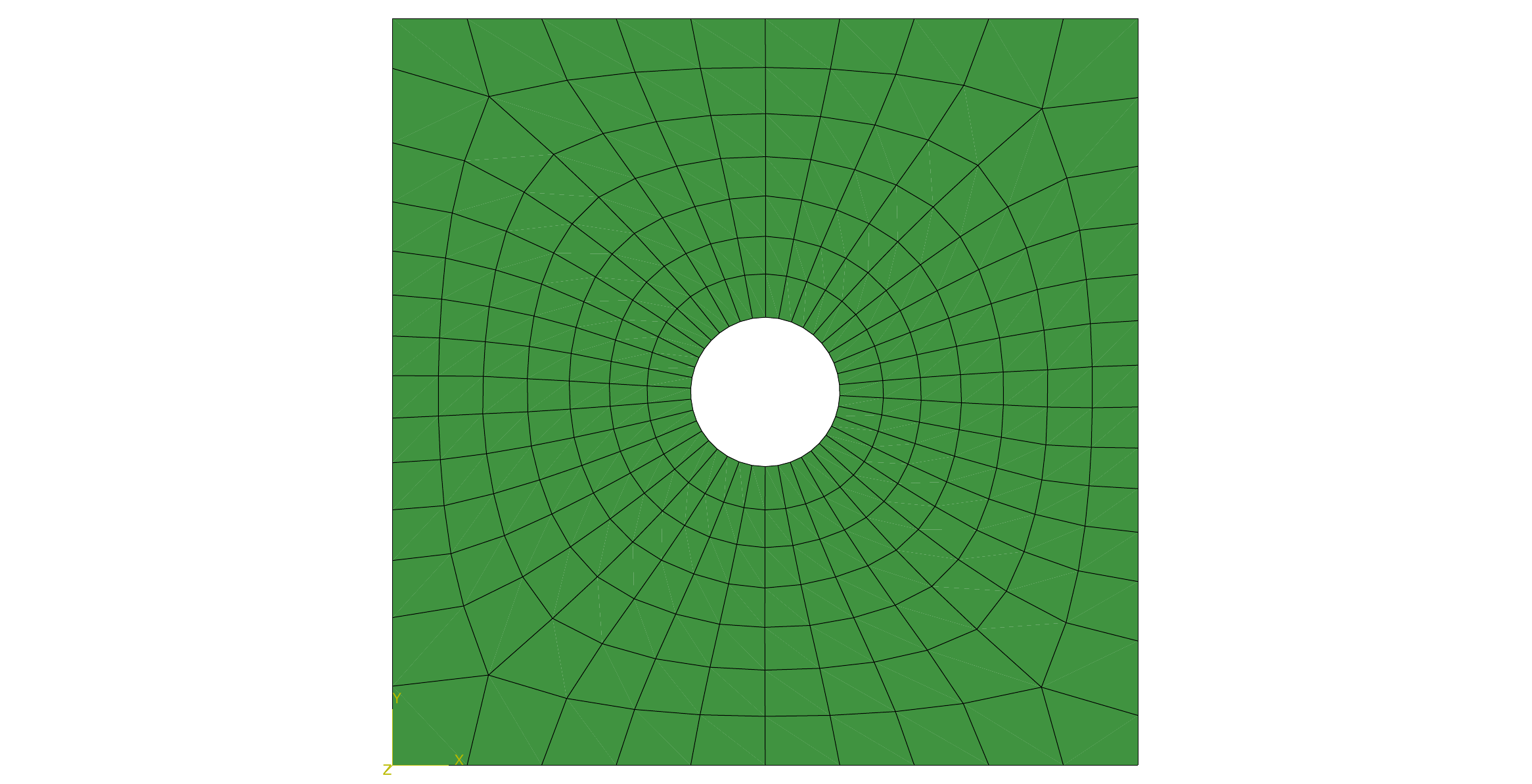}
            \caption{Finite element mesh of the composite plate.}
            \label{fig:appa_mesh}
        \end{figure}
        The plate's hole has a 6 mm diameter and the in-plane dimensions are both 5 times the hole diameter. The ply material is IM7/8552 prepreg (carbon fibres and epoxy matrix) and each ply has  $t=0.125$ mm thickness. We considered the lamination sequence $\left[45/90/-45/0\right]_S$ for a total of 8 plies and 1 mm laminate thickness.

    \subsection{Details of the FE models}
        All finite element models were done using the Abaqus finite element code \cite{abaqus2022} and Python scripting was used to automatically generate a different FE models for each of the strain loading combinations \cite{TomGulikers2018AbaqusANN,boyangChen2022summerschool}.
        
        The meshed part is illustrated in \Cref{fig:appa_mesh}, which shows that a radial mesh was obtained by seeding the hole edge 4 times as much as the outer edges. Since no delaminations were expected due to the absence of ply blocks, the elements were chosen to be S4 shells elements of the Abaqus Standard Element Library \cite{abaqus-manual2022}, whose in-plane and bending behaviour are described by the classical lamination theory (CLT), once the stacking sequence and ply thicknesses are specified. 
        
        Damage initiation was modeled with the Hashin criterion, while damage evolution was represented in a smeared crack fashion. For this purpose, the cohesive law available in Abaqus \cite{abaqus-manual2022} was employed to model the stiffness degradation due to matrix and fiber tensile and compressive failure.

    \section{Support vector machines, kernel methods and kernel-target alignment}
    \label{sec:appb}
    \subsection{Primal SVM}

    The SVM is the linear decision model
    \begin{equation}
        y = \bm{w}^\top\bm{x} + b
        \label{eq:algo_lin_svm_df}
    \end{equation}
    which assigns labels through the sign function
    \begin{equation}
        \mathrm{sgn}(y) = \mathrm{sgn}\left( \bm{w}^\top\bm{x} + b \right).
        \label{eq:algo_lin_svm_sgn_func}
    \end{equation}
    In \Cref{eq:algo_lin_svm_df,eq:algo_lin_svm_sgn_func}, $\bm{w}$ is the vector normal to the decision hyperplane and $b$ is the intercept.
    
    The optimal hyperplane is found by maximizing the \emph{geometric margin} of the dataset, which can be proved to be
    \begin{equation}
        \gamma^* = \frac{1}{\norm{\bm{w}}}.
        \label{eq:algo_lin_svm_gm}
    \end{equation}

    By minimizing the squared norm $\norm{\bm{w}}^2$ one obtaines the primal optimization problem of the SVM,
    \begin{equation}
        \begin{aligned}
            \min_{\bm{w},\, b}&\quad \frac{1}{2}\norm{\bm{w}}^2\\
            \mathrm{s.t.}&\quad y^{(m)}\left( \bm{w}^\top\bm{x}^{(m)} + b \right) \geq 1\quad m=1,\,\dots,\, M,
        \end{aligned}
        \label{eq:algo_lin_svm_primal}
    \end{equation}
    where $m$ identifies the sample and $M$ is the total number of training samples.

    \Cref{eq:algo_lin_svm_primal} enforces exact separability, which can lead to overfitting. A way to improve generalization is the so-called \emph{soft margin SVM}, which modifies \Cref{eq:algo_lin_svm_primal} by introducing the constraints slack variables $\xi^{(m)}$ and the penalty constant $C$,
    \begin{equation}
        \begin{aligned}
            \min_{\bm{w},\, b}&\quad \frac{1}{2}\norm{\bm{w}}^2 + C\sum_{m=1}^{M}\xi^{(m)}\\
            \mathrm{s.t.}&\quad y^{(m)}\left( \bm{w}^\top\bm{\varepsilon}^{(m)} + b \right) \geq 1 - \xi^{(m)}\quad m=1,\,\dots,\, M\\
            &\quad \xi^{(m)}\geq 0.
        \end{aligned}
        \label{eq:algo_lin_svm_primal_soft}
    \end{equation}

\subsection{Dual SVM and kernels}
    By introducing the Lagrange multipliers $\alpha^{(m)}$ and $\beta^{(m)}$, one can write the Lagrangian of the SVM optimization problem,
    \begin{equation}
        \begin{aligned}
            \mathcal{L}\left( \bm{w}, b, \bm{\xi}, \bm{\alpha} \right) =& \frac{1}{2}\ipc{\bm{w}}{\bm{w}} + C\sum_{m=1}^{M}\xi^{(m)} \\&- \sum_{m=1}^{M} \alpha^{(m)}\left( y^{(m)}\left( \bm{w}^\top\bm{x}^{(m)} + b \right) - 1 + \xi^{(m)} \right) - \sum_{m=1}^{M} \beta^{(m)} \xi^{(m)}.
            \label{eq:algo_svm_soft_lagrangian}
        \end{aligned}
    \end{equation}
    
    The \emph{dual} soft-margin SVM is obtained by setting all the derivatives of the Lagrangian in \Cref{eq:algo_svm_soft_lagrangian} equal to zero,
    \begin{equation}
        \begin{aligned}
            \max_{\bm{\alpha}}&\quad \sum_{m=1}^{M}\alpha^{(m)}-\frac{1}{2}\sum_{m,m^\prime=1}^{M}y^{(m)}y^{(m^\prime)}\alpha^{(m)}\alpha^{(m^\prime)}\ipc{\bm{x}^{(m)}}{\bm{x}^{(m^\prime)}}\\
            \mathrm{s.t.}&\quad 0\leq\alpha^{(m)}+\beta^{(m)}\leq C,\quad m=1,\,\dots,\,M\\
            &\quad\sum_{m=1}^{M}\alpha^{(m)}y^{(m)}=0.
    \end{aligned}
        \label{eq:algo_svm_soft_dual}
    \end{equation}

    \Cref{eq:algo_svm_soft_dual} is still a linear model in the original feature space. However, by introducing a feature map
    \begin{equation}
        \phi:\bm{x}\longrightarrow\phi(\bm{x})
    \end{equation}
    we can map the features nonlinearly and potentially to a manifold where they are more easily separable. Furthermore, replacing $\bm{x}$ with $\phi(\bm{x})$ in \Cref{eq:algo_svm_soft_dual}, we see that the mapped features only appear in the inner product
    \begin{equation}
        \ker{\bm{x}^{(m)}}{\bm{x}^{(m^\prime)}} = \ipc{\phi(\bm{x}^{(m)})}{\phi(\bm{x}^{(m^\prime)})},
        \label{eq:algo_kernel_general}
    \end{equation}
    which is known as the \emph{kernel} of the feature map. The advantage of having only inner product of features (\emph{kernel trick}) is the possibility of classifying in nonlinear feature spaces without having to compute the feature map explicitly.

    The kernels mostly used in machine learning are the polynomial, Gaussian and sigmoid kernels
    \begin{equation}
        \ker{\bm{x}}{\bm{x}^\prime}=
        \begin{cases}
            (\gamma\bm{x}^\top\bm{x}^\prime + c_0)^d&\quad \mathrm{(polynomial)}\\
            \exp\left( -\gamma\norm{\bm{x}-\bm{x}^\prime}^2 \right)&\quad \mathrm{(Gaussian)}\\
            \tanh\left( \gamma\bm{x}^\top\bm{x}^\prime + c_0 \right)&\quad \mathrm{(sigmoid)}
        \end{cases}
        \label{eq:algo_standard_kernels}
    \end{equation}
    
    \subsection{Kernel-target alignment}
    The \emph{alignment} between two kernels is defined as 
    \begin{equation}
        \begin{aligned}
            A\left( K^{(1)},\,K^{(2)} \right)&=\frac{\ipcf{K^{(1)}}{K^{(2)}}}{\sqrt{\ipcf{K^{(1)}}{K^{(1)}}\ipcf{K^{(2)}}{K^{(2)}}}}\\&=\frac{\ipcf{K^{(1)}}{K^{(2)}}}{\normf{K^{(1)}}\normf{K^{(2)}}},
        \end{aligned}
        \label{eq:appb_ka}
    \end{equation}
    where $K$ is the \emph{kernel matrix}, obtained by taking the kernel of all pairs of features, and
    \begin{equation*}
        \ipcf{K^{(1)}}{K^{(2)}}=\tr\left(K^{(1)\top} K^{(2)}\right).
    \end{equation*}
    The alignment between two kernels is always lesser or equal to 1, where 1 corresponds to perfect alignment.
    
    Assume a kernel $\kappa_{\bm{\theta}}$, parametrized by $\bm{\theta}$ and define the \emph{target kernel} matrix as
    \begin{equation}
        K^*=\mathbf{y}\mathbf{y}^{\top}.
        \label{eq:appb_kt}
    \end{equation}
    The \emph{kernel-target alignment} (KTA) of $\kappa_{\bm{\theta}}$ is the alignment between the chosen kernel and the target,
    \begin{equation}
        \begin{aligned}
            A\left( K_{\bm{\theta}},\,K^{*} \right)&=\frac{\ipcf{K_{\bm{\theta}}}{K^{*}}}{\normf{K_{\bm{\theta}}}\normf{K^{*}}}\\
            &=\frac{\ipcf{K_{\bm{\theta}}}{K^{*}}}{M\normf{K_{\bm{\theta}}}},
        \end{aligned}
        \label{eq:appb_kta}
    \end{equation}
    where $K_{\bm{\theta}}$ is the kernel matrix of $\kappa_{\bm{\theta}}$.

    The KTA enjoys theoretical properties such as concentration around its expected value and generalisation \cite{wang2015KtaReview} and therefore it is indicative of the ability of a kernel to separate classes of data.

    \section{Quantum computing notions}
    \label{sec:appc}

    \subsection{Quantum states}
    The basic logical unit in quantum computing is the \emph{qubit}. Mathematically speaking, this is a unit-norm vector in the complex 2-dimensional space $\mathbb{C}^{2}$ defined as a linear combination of two orthogonal basis states, $\ket{0}$ and $\ket{1}$, i.e.
    \begin{equation}
        \ket{\psi} = \psi_0\ket{0} + \psi_1\ket{1},\quad \psi_0,\,\psi_1\in\mathbb{C},
        \label{eq:appc_qubit}
    \end{equation}
    where the $\ket{\cdot}$ notation is used to indicate unit vectors. 
    
    As opposed to classical bits, \Cref{eq:appc_qubit} shows that a single qubit can be in any complex superposition of the two basis states. However, reading of a quantum state can only happen through a measurement, which will make the qubit collapse to one of the two basis states, $\ket{0}$ or $\ket{1}$. More specificallly, the qubit is measured as $\ket{0}$ with probability $p_0=\psi_0^2$ and as $\ket{1}$ with probability $p_1=\psi_1^2$. Since these are the only two possible outcomes, it must be that $p_0+p_1=\psi_0^2+\psi_1^2=1$, which explains the unitary norm of the qubit.

    Similarly, a states of $n$ qubits is defined as a superpositions of basis states that correspond to bitstrings, that is
    \begin{equation}
        \ket{\psi} = \psi_0\ket{0\dots 0}+\psi_1\ket{0\dots 1}+\dots+\psi_{N-1}\ket{1\dots 1}\quad \psi_0,\,\dots,\,\psi_{N-1}\in\mathbb{C},
        \label{eq:appc_multi_qubit}
    \end{equation}
    where $N=2^n$.

    The exponential relation between the number of qubits and the number of possible bitstrings speaks for the potential advantage of \emph{quantum superposition}, which allows multiple classical information states to be processed simultaneously through a quantum algorithm. Quantum superposition is at the heart of fundamental algorithms with proved complexity improvement such as quantum integer factoring \cite{Shor94} and quantum database search \cite{Grover96}.

    Nevertheless, the quantum state is inaccessible as readable information and measurement will collapse the wavefunction to only one of the $2^n$ basis states. Similarly to the single-qubit case, the basis state $\ket{i}$ has probability $p_i=\psi_i^2$ of being measured and
    \begin{equation}
        \sum_{i=0}^{N-1} p_i = \sum_{i=0}^{N-1} \psi_i^2 = 1.
        \label{eq:appc_unit_proba}
    \end{equation}

    Quantum states can be prepared by applying unitary transformations to a reference state, such as the all-zero state,
    \begin{equation}
        \ket{\psi}=U\ket{0}^{\otimes n},
        \label{eq:appc_generic_state_prep}
    \end{equation}
    where $U$ is the generic unitary transformation.

    \begin{figure}[tbp]
        \centering
        \includegraphics[width=0.4\textwidth]{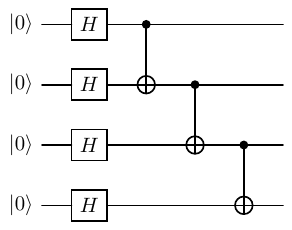}
        \caption{As an example of quantum circuit, the GHZ state can be prepared by first creating a uniform superpositon of all qubits through the Hadamard gate $H$ and then introducing entanglement with CNOT gates in cascade.}
        \label{fig:appc_ghz}
    \end{figure}

    \Cref{fig:appc_ghz} shows a unitary operation as a quantum circuit, i.e. a sequence of single- and two-qubit operations. Here, Hadamard gates $H$ are first applied to every qubit, where
    \begin{equation}
        \begin{aligned}
            H\ket{0} &= \frac{1}{\sqrt{2}}\left( \ket{0} + \ket{1} \right)\\
            H\ket{1} &= \frac{1}{\sqrt{2}}\left( \ket{0} - \ket{1} \right).
        \end{aligned}
        \label{eq:algo_hadamard}
    \end{equation}
    This first layer of Hadamard gates creates the uniform superposition state    \begin{equation}
        \frac{1}{\sqrt{2^n}}\left( \ket{0\dots 0},\,\ket{0\dots 1},\,\dots,\,\ket{1\dots 1} \right),
        \label{eq:algo_unif_sup_state}
    \end{equation}
    where each basis state can be sampled with the equal probability $1/2^n$. This is often a starting point state in many quantum algorithms.

    Following, CNOT gates act between neighbouring couples of qubits as
    \begin{equation}
        \begin{aligned}
            \mathrm{CNOT}\ket{00} &= \ket{00}\\
            \mathrm{CNOT}\ket{01} &= \ket{01}\\
            \mathrm{CNOT}\ket{10} &= \ket{11}\\
            \mathrm{CNOT}\ket{11} &= \ket{10}.
        \end{aligned}
    \label{eq:algo_cnot}
    \end{equation}
    CNOT gates are used to set qubits in an \emph{entangled} state, a condition in which any operation on any of the qubits affects also the rest of the state. In particular, the series of CNOT gates creates one of the maximally entangled Greenberger-Horne-Zeilinger (GHZ) states \cite{greenberger2007going}, specifically
    \begin{equation}
        \ket{\mathrm{GHZ}} = \frac{1}{\sqrt{2}}\left( \ket{0000}+\ket{1111} \right).
    \end{equation}
    
\subsection{Quantum embedding}
    State preparation can be used to embed classical data into quantum states, by mapping the features to a unitary transformation.
    \begin{equation}
        \ket{\phi\left( \bm{x} \right)}=U( \bm{x} )\ket{0},
        \label{eq:appc_qemb}
    \end{equation}
    A complete review of the different types of quantum embeddings is beyond the current scope and the interested reader is pointed to \cite{SchuldPetruccione2021} for a critical overeview.
    
\subsection{Quantum kernels}
    Quantum embeddings are effectively feature maps in the Hilbert space $\mathcal{H}\subseteq \mathbb{C}^{2^n}$. The kernel associated with it computes the overlap between quantum feature vectors in $\mathcal{H}$, that is
    \begin{equation}
        \kappa\left( \bm{\varepsilon}, \bm{\varepsilon}^\prime \right)=\abs{\ip{\phi\left( \bm{\varepsilon} \right)}{\phi\left( \bm{\varepsilon}^\prime \right)}}^2,
        \label{eq:appc_qkern}
    \end{equation}
    where the \emph{braket} notation $\ip{\cdot}{\cdot}$ indicates the inner product between two vectors in $\mathcal{H}$.

    By introducing \Cref{eq:appc_qemb} in \Cref{eq:appc_qkern}, the quantum kernel can be rewritten as
    \begin{equation}
        \kappa\left( \bm{x},\,\bm{x}^\prime \right)=\mel{0}{U^\dagger\left( \bm{x}^\prime \right)U\left( \bm{x} \right)}{0},
        \label{eq:appc_qkern_explicit}
    \end{equation}
    which shows that the quantum kernel can be computed as the probability of the all-zeros state, after applying the direct embedding for $\bm{x}$ and the reversed embedding for $\bm{x}^\prime$.

    \section{Classical and quantum SVM comparison on different classification metrics}
    \label{sec:appd}
    \begin{table}[h]
        \centering
        \begin{tabular}{cccccc}
            \hline
            $N_{\mathrm{train}}$ & Accuracy & Jaccard Index & Precision & Recall & Specificity \\ \hline
            \multicolumn{6}{c}{RBF kernel}                                                     \\ \hline
            156                  & 0.694    & 0.627         & 0.784     & 0.759  & 0.795       \\
            313                  & 0.750    & 0.708         & 0.835     & 0.824  & 0.838       \\
            470                  & 0.788    & 0.751         & 0.886     & 0.832  & 0.895       \\
            627                  & 0.788    & 0.750         & 0.893     & 0.825  & 0.903       \\
            784                  & 0.838    & 0.813         & 0.939     & 0.859  & 0.944       \\
            940                  & 0.824    & 0.797         & 0.915     & 0.861  & 0.921       \\
            1097                 & 0.848    & 0.827         & 0.938     & 0.874  & 0.943       \\
            1254                 & 0.878    & 0.866         & 0.943     & 0.913  & 0.945       \\
            1411                 & 0.873    & 0.860         & 0.937     & 0.912  & 0.939       \\
            1568                 & 0.882    & 0.869         & 0.955     & 0.906  & 0.958       \\ \hline
            \multicolumn{6}{c}{HE2W6D3 kernel}                                                 \\ \hline
            156                  & 0.699    & 0.626         & 0.804     & 0.739  & 0.823       \\
            313                  & 0.731    & 0.675         & 0.835     & 0.780  & 0.846       \\
            470                  & 0.756    & 0.708         & 0.858     & 0.802  & 0.869       \\
            627                  & 0.779    & 0.741         & 0.878     & 0.826  & 0.887       \\
            784                  & 0.795    & 0.762         & 0.892     & 0.839  & 0.900       \\
            940                  & 0.794    & 0.757         & 0.897     & 0.829  & 0.907       \\
            1097                 & 0.805    & 0.773         & 0.902     & 0.844  & 0.910       \\
            1254                 & 0.797    & 0.765         & 0.884     & 0.849  & 0.891       \\
            1411                 & 0.814    & 0.785         & 0.911     & 0.851  & 0.917       \\
            1568                 & 0.818    & 0.790         & 0.912     & 0.855  & 0.919       \\ \hline
        \end{tabular}
        \vspace{1em}
        \caption{Comparison of five different classification scores between the RBF-SVM and the trained HE2W6D3 quantum kernel SVM. The classifiers were trained with increasing fractions of the training dataset. Notice that the RBF-SVM problem used $C=10^7$, while the HE2W6D3 quantum kernel SVM used $C=10^4$, which is the highest $C$ values before the occurrence of convergence issues.}
    \end{table}

    \bibliographystyle{unsrt}
    \bibliography{references}

\end{document}